\documentclass[10pt,letterpaper,amsmath,amsfonts,amssymb,aps,pra,reprint,superscriptaddress]{revtex4-1}
\usepackage[latin1]{inputenc}
\usepackage{bm}
\usepackage[pdftex]{graphicx}

\newcommand{\op}[1]{\hat{\bm #1}}
\newcommand{\ket}[1]{\lvert #1\rangle}
\newcommand{\bra}[1]{\langle #1 \rvert}
\newcommand{\pr}[1]{\ket{#1}\bra{#1}}
\newcommand{\ipr}[2]{\langle #1 \vert #2 \rangle}
\newcommand{\mean}[1]{\left\langle #1 \right\rangle}
\newcommand{\cw}{\circlearrowright}
\newcommand{\ccw}{\circlearrowleft}

\begin{document}
\title{Strengthening weak value amplification with recycled photons}
\author{Justin Dressel}
\author{Kevin Lyons}
\affiliation{Department of Physics and Astronomy and Rochester Theory Center, University of Rochester, Rochester, New York 14627, USA}
\author{Andrew N. Jordan}
\affiliation{Department of Physics and Astronomy and Rochester Theory Center, University of Rochester, Rochester, New York 14627, USA}
\affiliation{Institute of Quantum Studies, Chapman University, 1 University Drive, Orange, CA 92866, USA}
\author{Trent M. Graham}
\author{Paul G. Kwiat}
\affiliation{Department of Physics, University of Illinois at Urbana-Champaign, Urbana, Illinois 61801-3080, USA}

\date{\today}

\begin{abstract}
  We consider the use of cyclic weak measurements to improve the sensitivity of weak-value amplification precision measurement schemes.  Previous weak-value experiments have used only a small fraction of events, while discarding the rest through the process of ``post-selection''.  We extend this idea by considering recycling of events which are typically unused in a weak measurement.  Here we treat a sequence of polarized laser pulses effectively trapped inside an interferometer using a Pockels cell and polarization optics. In principle, all photons can be post-selected, which will improve the measurement sensitivity.  We first provide a qualitative argument for the expected improvements from recycling photons, followed by the exact result for the recycling of collimated beam pulses, and numerical calculations for diverging beams.  We show that beam degradation effects can be mitigated via profile flipping or Zeno reshaping.  The main advantage of such a recycling scheme is an effective power increase, while maintaining an amplified deflection. 
\end{abstract}

\maketitle

\section{Introduction}
A long standing goal in optics is the development and improvement of precision optical metrology.  In the first paper on weak values in 1988 \cite{Aharonov1988,*Duck1989}, Aharanov, Albert and Vaidman suggested that the weak value effect might be used as an amplifier in order to measure (in the case they were considering) the value of a small magnetic field by looking at the anomalously large deflection of a beam of atoms traversing a Stern-Gerlach apparatus.  The general validity of this weak value effect was later shown experimentally in an optical context by Ritchie \emph{et al.} \cite{Ritchie1991}, who replaced the magnetic spin with transverse polarization, and Brunner \emph{et al.} \cite{Brunner2003}, who illustrated the pervasiveness of the weak value effect in common optical telecom networks.

More recently, the amplification properties of this weak value effect have been exploited in similar optical systems to precisely measure beam deflection \cite{Hosten2008,Dixon2009,Starling2009,Turner2011,Pfeifer2011,Hogan2011,Zhou2012}, phase shifts \cite{Starling2010b,*Starling2010a}, frequency shifts \cite{Starling2010}, time delays \cite{Strubi2013}, and even temperature shifts \cite{Egan2012}, by using either polarization or which-path degrees of freedom.  Although these experiments can be described using classical wave optics \cite{Howell2010}, the analysis using quantum techniques provides additional insight and allows for future extension to cases with no classical counterpart.  Hence, we shall continue to use a quantum approach in this work as well.

Our theoretical analysis begins with the Rochester setup \cite{Dixon2009}, where the tilt of a moving mirror within an interferometer is detected from the signal on a split-detector.  While this setup has a sub-picoradian resolution with only milliwatts of laser power, there are a number of ways this can be improved to yield even greater sensitivity.

A generic shortcoming of weak-value-related metrological techniques is the fact that only a small fraction of the events are ``post-selected'', while the vast majority of events are intentionally thrown away.  The main goal of the current work is to investigate how this situation can be further improved if those events are recycled.  This will be done by taking photons which are not post-selected and reinjecting them back into the interferometer, so that eventually, every photon can be post-selected in principle.  We will see that this strategy does indeed lead to an improvement in the signal-to-noise ratio of the desired parameter, effectively given by the power increase of the split-detection signal.  Moreover, since the existing single-pass weak-value amplification already achieves the sensitivity of standard measurement techniques (such as homodyne detection) but with lower technical noise \cite{Starling2009}, the improvements from recycling should \emph{exceed} the sensitivity of the standard techniques.  We note that because we employ not just a single pass, but many passes of a given photon through the interferometer, the simple weak value formula used in the first paper on the subject \cite{Dixon2009} will no longer suffice, and we must develop a theoretical formalism for multiple passes that will account for the amplification of the deflection, as well as the probability of reaching the detector after some number of traversals.

While the recycling scheme is an important advance in its own right (and can be generically applied to all weak value amplification schemes), it also lends itself to further enhancement if combined with other precision metrology techniques currently in use.  For example, the inclusion of a spatial filter or parity-flipping element to Zeno-stabilize the beam, or the use of a squeezed reference beam \cite{Caves1981,Barnett2003,Treps2002,Treps2003} could significantly reduce degradation effects and quantum noise, respectively.  The recycling technique, therefore, sets the stage for combined weak value/quantum light amplification strategies for future research.  Furthermore, though our present work focuses on a novel pulsed recycling method, possible extensions to continuous wave operation may allow for the use of power recycling \cite{Drever1983} and signal recycling \cite{Meers1991} techniques, both of which are in use in modern gravitational wave detectors \cite{Schnier1997,Vahlbruch2005}. 

This paper is organized as follows.  In Section~\ref{sec:qual} we give a heuristic estimation of the expected gains from a recycling setup based on qualitative power considerations and accessible laboratory conditions.  In Section~\ref{sec:collimated} we analytically compute the exact recycling solution for a particular setup under the assumption that the beam can stay perfectly collimated.  In Section~\ref{sec:diverging} we relax this assumption with a numerical treatment including diffraction effects.  We summarize our conclusions in Section~\ref{sec:conclusion}.

\section{Qualitative Arguments}\label{sec:qual}
\begin{figure}[t]
  \begin{center}
    \includegraphics[width=\columnwidth]{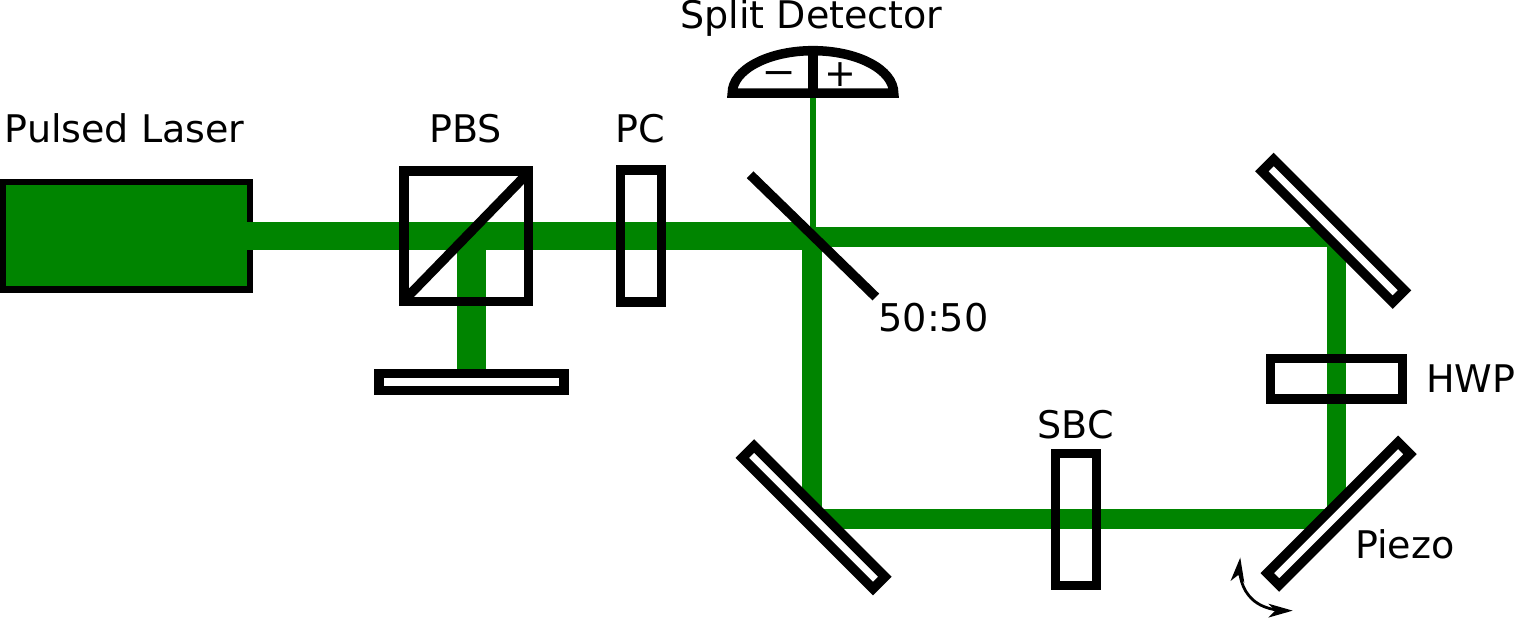}
  \end{center}
  \caption{Simple cyclic weak measurement scheme.  A laser emits a pulse of horizontal (H) polarization through a polarizing beam splitter (PBS), which travels through an active Pockels cell (PC) that rotates the polarization to vertical (V), after which it enters a Sagnac interferometer through a 50:50 beam splitter.  Inside the interferometer the combination of half-wave plate (HWP) and Soleil-Babinet compensator (SBC) rotates the pulse polarization back to H with a relative tunable phase shift of $\phi$ between the clockwise ($\cw$) and counter-clockwise ($\ccw$) traveling paths.  The piezo-driven mirror imparts a transverse momentum kick $k$ that differs by a sign for the $\cw$ and $\ccw$ paths.  A split detector is placed at the dark port to measure a resulting pulse deflection.  The H-polarized part of the pulse that exits the bright port is rotated again by the active PC back to V before being confined by the PBS and mirror to return the pulse to the interferometer through the PC, now \emph{inactive}.}
  \label{fig:setup}
\end{figure}

Our baseline for comparison will be the Rochester continuous wave (CW) Sagnac interferometric scheme described in \cite{Dixon2009,Starling2009,Howell2010}.  We wish to improve the detected signal-to-noise ratio (SNR) by using a combination of pulsed laser operation with the same average power output and a design that sends the undetected portion of each pulse back into the interferometer.  Such a setup is illustrated in Figure~\ref{fig:setup} for reference.  However, before committing to a particular recycling design we can make fairly general estimations about the increases in sensitivity that we expect from any similar recycling scheme.  

\subsection{Characteristic Time Scales}
The constraints on how much we can increase the power collected by the dark port detector in order to improve the measurement sensitivity depend crucially on the relative time scales involved, which include:
\begin{enumerate}
  \item The pulse duration $\tau$ being emitted by the laser.  For typical lasers this can vary between $1$ ns and $5$~fs reasonably, which correspond to pulse lengths of $0.3$ m and $1.5~\mu$m, respectively.
  \item The repetition period $T = 1/f$ of the laser.  For typical lasers the repetition rate $f$ can vary from $1$~Hz to several GHz reasonably.
  \item The traversal period $T_r$ of the interferometer setup.  This is determined by the physical size of the setup.  As an upper-bound estimate, a $3$ m long recycling setup will have a total period of $10$ ns.
  \item The gating time $T_g$ for adding new pulses to the interferometer.  This will determine the minimum inter-pulse spacing $T_p = \tau + T_g$ inside the interferometer.  This also must be strictly less than the time between pulses emitted by the laser $T_g < T - \tau$ so that every new laser pulse can be injected.  For a Pockels cell, $T_g \approx 2$ ns.
\end{enumerate}
These are summarized in Table~\ref{tab:timescales}.

\begin{table}
  \centering
  \begin{tabular}{l | l r}
    Symbol \; & \; Description & Estimate \\
    \hline
    $\tau$ & \; Laser pulse duration & 5 fs - 1 ns \\
    $T$ & \; Laser repetition period & 1 ns - 1 s \\
    $T_r$ & \; Traversal period & 1 ns - 10 ns \\
    $T_g$ & \; Gating time & 2 ns \\
    $T_p$ & \; Minimum inter-pulse spacing & $\tau + T_g$
  \end{tabular}
  \caption{Relevant time scales for a recycling experiment.}
  \label{tab:timescales}
\end{table}

We assume in what follows that $\tau < T_p < T_r$, so that at least one pulse can be trapped inside the interferometer.  We also assume that the average power output $P$ of the pulsed laser is equal to the average power of a comparison CW beam.  As a result, the average power of each individual pulse will be increased by a factor $T/\tau$ from the reference CW beam.

\subsection{Detector SNR}\label{sec:snr}
The SNR is a useful indicator for the sensitivity of the measurement, since a signal producing an SNR of unity indicates the smallest practically resolvable signal \cite{Barnett2003}.  The detected SNR is defined as the ratio of the collected signal to the square root of the variance of that collected signal.  Our raw signal $\mean{S}$ is a split-detection of the transverse profile of the beam, which measures the difference in photon number collected by each side of the detector, thereby providing information about the horizontal displacement of the beam.  For small displacements, the variance of the split-detected signal is well approximated by the second moment, which is in turn proportional to the total photon number for position-uncorrelated photons (see, for example, Ref~\cite{Barnett2003}).

The total accumulated split-detected signal scales linearly with the average collected energy, which can be factored into the average power at the detector $P_d$ multiplying the collection duration $t$.  The variance will be similarly proportional to $P_d t$, so the SNR will scale as $P_dt/\sqrt{P_dt} = \sqrt{P_dt}$.  Hence, the SNR can be increased either by waiting for a longer duration $t$ or by increasing the average power $P_d$ at the dark port detector.  Our recycling scheme proposes to increase the average power collected within a fixed duration to increase the sensitivity.  

In the original interferometric weak value scheme, the detector collected a power of $P_d= \eta P$, where $\eta$ was the post-selected fraction of the total laser power $P$ coming from the dark port. If we recycle the unused light, however, the average power $P_d$ collected at the detector after $r_T$ recycling passes in a laser repetition period $T$ will have the modified form, 
\begin{align}\label{eq:power}
  P_d &= \sum_{n=1}^{r_T} (1-\eta)^{n-1} \eta P = (1-(1-\eta)^{r_T}) P,
\end{align}
where $\eta$ is the fraction of the input power that exits the dark port of the interferometer after each traversal, and $r_T$ is the number of recycled pulses that hit the detector.  Here we have ignored optical losses and detector inefficiencies for clarity.  The power collected at the detector after a single traversal is $\eta P$ and the SNR scales as $\sqrt{P_d}$, so the net SNR gain factor will be,
\begin{align}\label{eq:powergain}
  \sqrt{\frac{P_d}{\eta P}} &= \sqrt{\frac{1 - (1-\eta)^{r_T}}{\eta}}.
\end{align}

For a small post-selection probability---such as those used in weak value experiments---then we can expand \eqref{eq:powergain} around $\eta = 0$ to find,
\begin{align}
  \sqrt{\frac{P_d}{\eta P}} &\approx \sqrt{r_T}\left(1 - (r_T-1) \frac{\eta}{4}\right) + O(\eta^2).
\end{align}
For $\eta (r_T-1) \ll 1$, then we can neglect the attenuation of the pulse to see an approximate $\sqrt{r_T}$ SNR scaling.

For sufficiently large $r_T$, however, the factor \eqref{eq:powergain} saturates to the constant value $\sqrt{1/\eta}$.  This saturation stems from the progressive attenuation of the recycled pulse.  Furthermore, the smaller the post-selection probability gets, the larger we can make the possible SNR gain over a single pass.  In this limit, however, $P_d \to P$ according to \eqref{eq:power} and all the photons will be collected.  Note that despite the large gain in power \eqref{eq:powergain} at the detector, the best SNR that one can obtain still scales according to the standard quantum limit.

The measured signal at the detector may be additionally modified by geometric and propagation effects, which we can encapsulate by an overall factor $\xi(r_T)$ that depends on $r_T$.  The total SNR gain factor over an unrecycled pulse will then be $\xi(r_T)\sqrt{P_d/\eta P}$.  For the sake of comparison, we will initially ignore these effects on the signal, so we will approximate $\xi(r_T) \approx 1$ in our qualitative arguments.  We will see in Section \ref{sec:collimated} that for a collimated beam $\xi(r_T)$ will approximate unity for small $r_T$ but will eventually converge to zero for large $r_T$.  Corrections to this effect will be discussed in Sections \ref{sec:parity} and \ref{sec:zeno}, where we will see that one can maintain a measurable signal for a collimated beam by inverting photons about the optical axis on each traversal or Zeno-stabilizing the beam with a spatial filter.  We shall also see in Section \ref{sec:diverging} that $\xi(r_T)$ can exceed unity for a carefully chosen pulse divergence, which can compensate for the attenuation effects and recover the approximate $\sqrt{r_T}$ scaling for a much wider range of $r_T$.

\subsection{Recycled Pulse Number}\label{sec:pulsenumber}
We can compute the number of pulses $r_T$ that hit the detector per laser repetition period $T$ from two factors.  First, each trapped pulse can traverse the interferometer roughly $r \le T / T_r$ times each repetition period.  Each traversal contributes one additional pulse impact to the detector.  Second, one can accumulate a maximum of $p \le T_r / T_p = T_r / (\tau + T_g)$ pulses that are trapped inside the interferometer.  Hence the total number of detector impacts $r_T = p r$ per period $T$ will be bounded by $T / (\tau + T_g)$.  Correspondingly, the maximum SNR gain factor \eqref{eq:powergain} that we can expect from power considerations will also be bounded entirely by the relative time scales and the post-selection probability.

In practice, not every recycled pulse will contribute constructively to the SNR.  Indeed, as shown in \eqref{eq:powergain} and as we shall see in Section~\ref{sec:collimated}, there will be some maximum number $r_{\text{max}}$ of constructive pulse impacts before the SNR saturates or decays.  To maximize the SNR gain in such a case, the pulse should be discarded and replaced by a fresh pulse.  Hence, the number of practical detector collections $r_T$ will be less than the maximum estimation $r_T = p r_{\text{max}} \le T / (\tau + T_g)$, so the number of pulses $p$ that can fit inside the interferometer will become important.  Both pulse stabilization techniques and diverging lenses can increase the practical range of $r_{\text{max}}$, which we will discuss in Sections~\ref{sec:parity}, \ref{sec:zeno} and \ref{sec:diverging}.


\subsection{Practical Estimates}\label{sec:snrestimates}
Using the Pockels cell as a gate, we expect $T_g \approx 2$ ns.  Assuming a short pulse $\tau \ll T_g$, then the inter-pulse spacing will be $T_p c \approx T_g c = 0.6$ m.  It follows that the maximum number of pulses inside the interferometer will be $p \approx T_r / T_g$.  Assuming a large $3$m setup, $T_r \approx 10$ ns, so $p \approx 10 / 2 = 5$ will be a generous upper bound to the number of pulses that we can expect to fit inside any interferometer.  For contrast, the smallest setup that fits only a single pulse will be $T_r = T_p$, or $T_p c \approx 0.6$ m in length.

As shown in Section~\ref{sec:regimes}, without loss or stabilization we can expect $r_{\text{max}} \leq 80$ to be an optimistic upper bound for a constructive number of recycling passes.  The maximum number of pulse impacts $r_T = p r_{\text{max}}$ per period $T$ that we expect with the largest setup of $p=5$ pulses is thus $r_T \leq 400$.  Therefore, we can expect an SNR gain to span the range from a maximum of $\sqrt{400}=20$ over a single pass for very small post-selection probability $\eta$ to $\sqrt{1/\eta}$ for larger $\eta$ according to \eqref{eq:powergain}.  Since $r_{\text{max}} \le T / T_r$ and $T_r \approx 10$ ns for the $3$ m setup, the laser repetition period must be $T \approx 400$ ns, implying a repetition rate of $f \approx 2.5$ MHz.  For contrast, the smallest setup of $0.6$m can fit only $p = 1$ pulse, so $r_T \leq 80$.  The SNR gain thus ranges from a maximum of $\sqrt{80}\approx9$ over a single pass to $\sqrt{1/\eta}$.  The $0.6$ m setup has recycling period $T_r \approx 2$ns, so must have a laser repetition period $T \leq 160$ ns, or rate $f \geq 6.25$ MHz.  These laser specifications should be readily achievable in the laboratory.

\section{Analytic Results}\label{sec:collimated}
\subsection{Sagnac Interferometer}
Following the experiment described in \cite{Dixon2009,Starling2009,Howell2010}, we extend the schematic to pulsed laser operation and pulse recycling.  As shown in Fig.~\ref{fig:setup}, the addition of a Pockels cell (PC) and polarizing beam splitter (PBS) allows the unused portion of each pulse that exits the bright port of the Sagnac interferometer to be redirected back inside the interferometer to complete multiple traversals.  The combination of half-wave plate (HWP) and Soleil-Babinet compensator (SBC) provide a tunable relative phase $\phi$ between the clockwise ($\cw$) and counter-clockwise ($\ccw$) propagating paths of the interferometer, but also flips the net polarization of each pulse.  As a result, the PC must be active as each pulse initially enters the bright port and when each pulse exits the bright port again; however, it must be \emph{inactive} as each pulse returns to the bright port after being confined by the PBS and mirror.  By injecting new pulses exactly when older pulses exit the bright port, one can minimize the inter-pulse spacing inside the interferometer to roughly a single gating time.

We also briefly note that the HWP and SBC can be removed in favor of a vertical tilt to provide the relative phase $\phi$.  With this variation, the PC turns on and off only once per repetition period in order to inject a new pulse into the interferometer, as opposed to cycling for every pulse traversal.  This variation does not change the inter-pulse spacing inside the interferometer, however, so it provides no SNR benefits, though it does provide a technical advantage due to the minimized number of PC cycles per laser repetition period $T$.

\subsection{Pulse Recycling}
Because there is no important interaction between distinct pulses in the recycling scheme, the SNR gains are fundamentally determined by the effects of single pulse recycling.  Therefore, we shall consider in some detail what happens to a single pulse profile after $r$ passes through the interferometer under the assumption that the pulse remains collimated.  We will relax the collimation assumption numerically in Section \ref{sec:diverging}. 

\begin{figure*}[th]
  \begin{center}
    \includegraphics[width=6in]{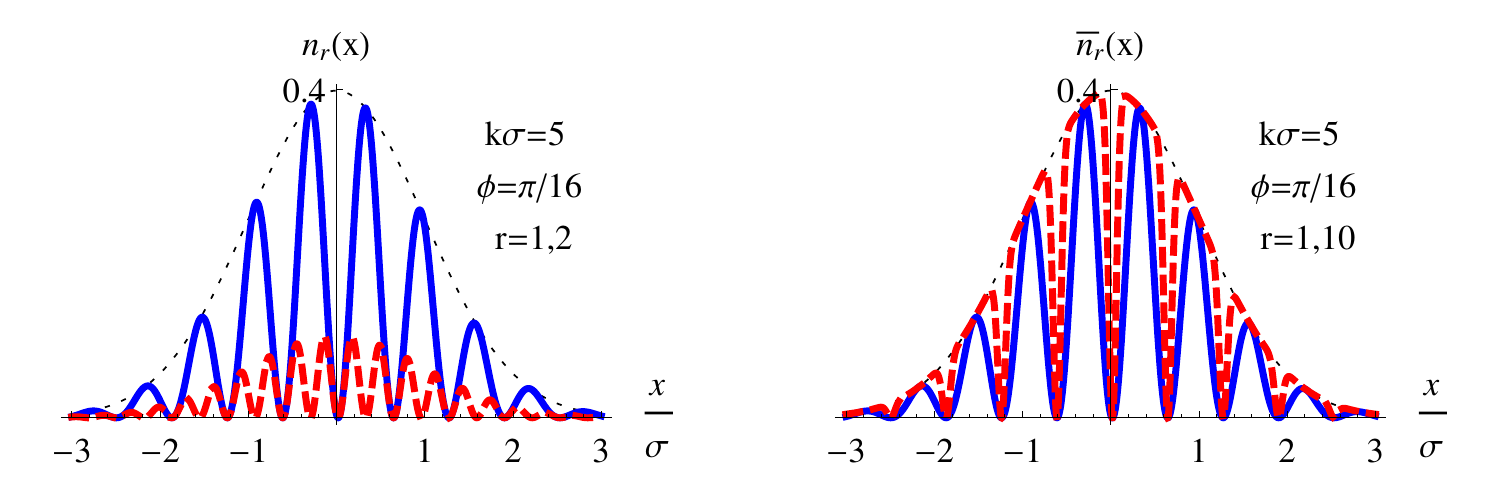}
  \end{center}
  \caption{Strongly misaligned regime with $\phi < 1 < k\sigma$.  Left: the transverse pulse profile $n_r(x)$ that impacts the dark port detector on the first (blue, solid) and second (red, dashed) traversals.  Right: the accumulated transverse pulse profile $\bar{n}_r(x)$ on the dark port detector after a single (blue, solid) and ten (red, dashed) traversals.  For this regime the interference pattern covers the entire profile, subsequent pulses are strongly attentuated, and the interference of the accumulated profile is slowly filled in. }
  \label{fig:beam1}
\end{figure*}

\subsubsection{Pulse States}
Assume the clockwise-propagating state of the pulse in the Sagnac interferometer is denoted $\ket{\cw}$ and the counter-clockwise-propagating state is denoted $\ket{\ccw}$.  Then the state that enters the interferometer through the 50:50 beam splitter will have the form, $\ket{\psi_+} = \frac{1}{\sqrt{2}}\left( \ket{\cw} + i \ket{\ccw} \right)$.  This will also be the post-selection state for the bright port of the interferometer.  Similarly, the post-selection state for the dark port of the interferometer will have the orthogonal form, $\ket{\psi_-} = \frac{1}{\sqrt{2}}\left( \ket{\cw} - i \ket{\ccw} \right)$.  We also define the which-path operator as $\op{W} = \pr{\cw} - \pr{\ccw}$.

Assume the initial transverse pulse profile is given by a state $\ket{\varphi}$.  We normalize the state of the transverse pulse profile so that its squared norm will encode the average photodetection rate.  Hence, measuring a pulse with a detector for the pulse duration $\tau$ will produce $N = \tau||\varphi||^2$ photon impacts upon the detector per pulse on average.  This choice of normalization will allow simple computation of the SNR without multi-particle Fock space calculations (e.g., as used in \cite{Barnett2003}).  

The total initial pulse state that enters the interferometer will have the product form, $\ket{\Psi_0} = \ket{\psi_+}\ket{\varphi}$.  For simplicity we suppress the polarization of the state and any propagation effects.

The traversal through the interferometer performs three operations on the state.  The first is the passage through the SBC and HWP, which creates a relative phase shift $\phi$ between the paths that can be described by the unitary operator $\op{U}_{\text{SBC}} = e^{i \phi \op{W} / 2}$.  The second is the tilting piezo mirror, which imparts a transverse momentum kick $k$ to the pulse, described by the unitary operator $\op{U}_{\text{P}} = e^{-i k \op{W}\op{x}}$, where $\hbar = 1$ and the transverse position operator $\op{x}$ generates a momentum translation $k$.  The third is a generic uniform loss with probability $\gamma$, described by a nonunitary loss operator, $\op{L} = \sqrt{1-\gamma}\op{1}$.  

The state of the pulse profile as it arrives back at the 50:50 beam splitter after one traversal has the form, $\ket{\Psi_1} = \op{L}\op{U}_{\text{P}}\op{U}_{\text{SBC}}\ket{\Psi_0}$.  After the pulse traverses the 50:50 beam splitter, it splits into two paths once more.  The dark port projects the photon onto the $\ket{\psi_-}$ state, and the bright port projects the photon onto the $\ket{\psi_+}$ state.  Hence, we obtain the following two states in the bright and dark ports, respectively, $\ket{\Psi_\pm} = \ket{\psi_\pm} \left(\op{M}_\pm \ket{\varphi} \right)$, where we have factored out the measurement operators $\op{M}_\pm = \bra{\psi_\pm} \op{L} \op{U}_{\text{P}} \op{U}_{\text{SBC}} \ket{\psi_+}$ that affect the transverse profile of the pulse in each case.  Written out explicitly, these measurement operators are diagonal in the position basis and have a remarkably simple form,
\begin{subequations}\label{eq:measops}
\begin{align}
  \op{M}_+ &= \sqrt{1-\gamma} \cos\left(\phi/2 - k\op{x}\right), \\
  \op{M}_- &= i \sqrt{1-\gamma} \sin\left(\phi/2 - k\op{x}\right), 
\end{align}
\end{subequations}
where we have used $\bra{\psi_\pm}\op{W}^n\ket{\psi_+} = (1 \pm (-1)^n)/2$.

\subsubsection{Number Densities}\label{sec:number}
Using the measurement operators \eqref{eq:measops}, the exact pulse state that exits the dark port after $r$ traversals through the interferometer will be $\ket{\Psi^r_-} = \ket{\psi_-} \left(\op{M}_- (\op{M}_+)^{r-1} \ket{\varphi}\right)$.  Therefore, the number density $n_r(x)$ of photons that hit the dark port detector at a transverse position $x$ on the $r$\textsuperscript{th} pulse traversal is,
\begin{align}\label{eq:nminus}
  n_r(x) &= \tau |\ipr{x}{\Psi^r_-}|^2 = n_0(x) (1-\gamma)^r \times \\
  &\qquad \sin^2\left(\frac{\phi}{2} - kx\right)\cos^{2(r-1)}\left(\frac{\phi}{2} - kx\right), \nonumber
\end{align}
where $n_0(x) = \tau |\ipr{x}{\varphi}|^2$ is the number density for the input pulse.

The total number density $\bar{n}_r(x)$ that accumulates on the dark port detector after $r$ traversals of the pulse will be the sum of the number densities for the $r$ traversals,
\begin{align}\label{eq:nminusacc}
  \bar{n}_r(x) &= \sum_{j=1}^r n_j(x) \\
  &= n_0(x) \frac{(1-\gamma)\left(1 - \left[(1-\gamma) \cos^{2}\left(\phi/2 - kx\right)\right]^r\right)}{1 + \gamma \cot^2\left(\phi/2 - kx\right)}. \nonumber
\end{align}
Hence the total number of photons that hit the detector after $r$ traversals is $N_r = \int \! \textrm{d}x\, \bar{n}_r(x)$.  Furthermore, if we compare \eqref{eq:nminusacc} to the heuristically estimated detector power \eqref{eq:power} when $\gamma\to 0$, we see that the spatially resolved version of the post-selection probability is given by $\eta \leftrightarrow \sin^2(\phi/2 - k x)$.

\begin{figure*}[th]
  \begin{center}
    \includegraphics[width=6in]{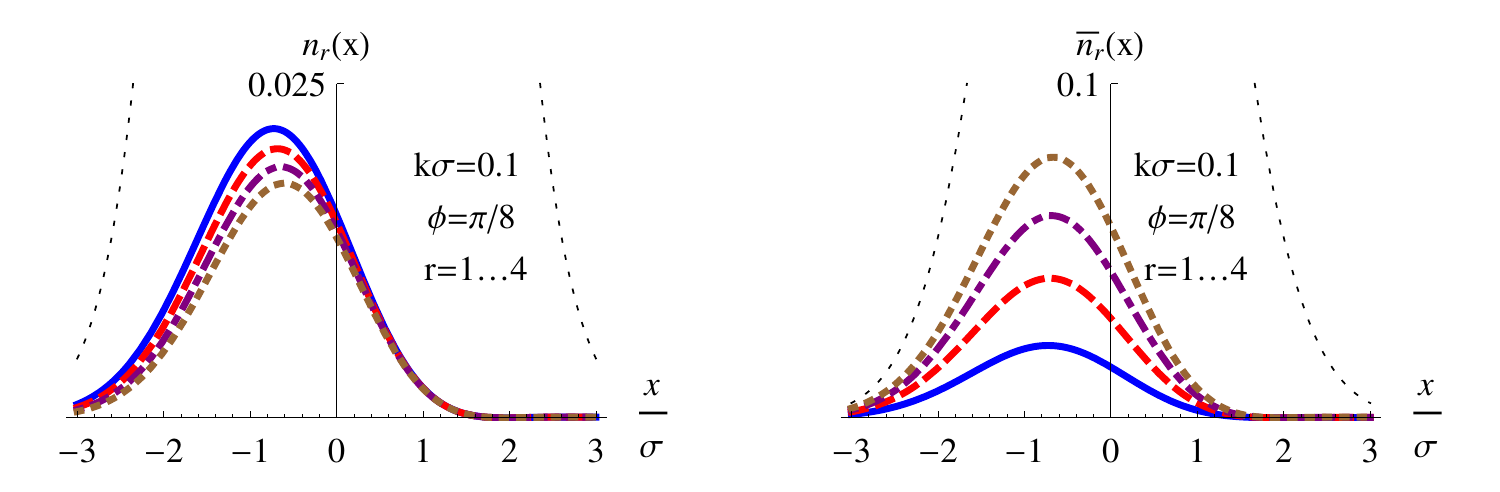}
  \end{center}
  \caption{Weak-value regime with $k\sigma < \phi < 1$, with parameters chosen to exaggerate the walk-off effect.  Left: the transverse single lobed pulse profile $n_r(x)$ that impacts the dark port detector on the first four traversals in order of (blue, solid), (red, dashed), (purple, dot-dashed), and (brown, dotted).  Right: the accumulated transverse pulse profile $\bar{n}_r(x)$ on the dark port detector after the first four traversals, with the same color coding.  For this regime, the dark port profile resembles a single shifted Gaussian that gradually walks back toward the center on multiple traversals with some attenuation, eventually recovering the original profile.  For more realistic parameters, such as those used in \cite{Dixon2009}, the walk-off effect is smaller per traversal. }
  \label{fig:beam2a}
\end{figure*}

\begin{figure*}[th]
  \begin{center}
    \includegraphics[width=6in]{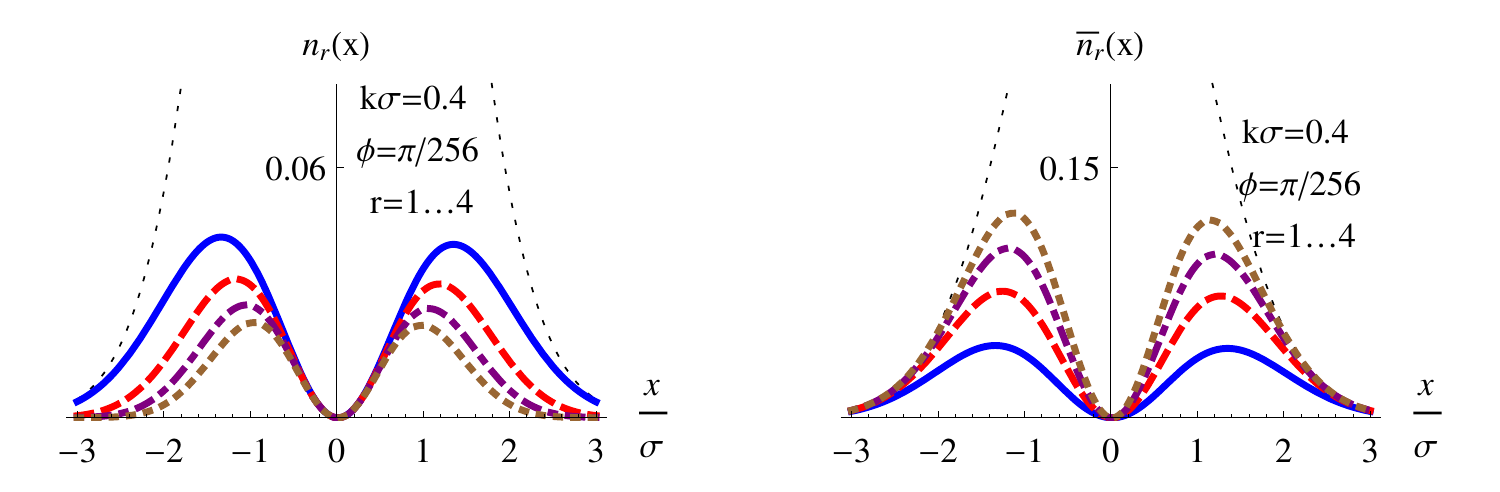}
  \end{center}
  \caption{Inverse weak-value regime with $\phi < k\sigma < 1$, with parameters chosen to exaggerate the walk-off effect.  Left: the transverse double-lobed pulse profile $n_r(x)$ that impacts the dark port detector on the first four traversals in order of (blue, solid), (red, dashed), (purple, dot-dashed), and (brown, dotted).  Right: the accumulated transverse pulse profile $\bar{n}_r(x)$ on the dark port detector after the first four traversals, with the same color coding.  For this regime, there are two lobes that very gradually walk back toward the center on multiple traversals with some attenuation to eventually recover the original beam profile.  For more realistic parameters, such as those used in \cite{Starling2010a,Starling2010b}, the walk-off effect is smaller per traversal. }
  \label{fig:beam3a}
\end{figure*}

In the limit of an infinite number of trials $r \to \infty$, the final term in \eqref{eq:nminusacc} vanishes and we are left with the number density,
\begin{align}\label{eq:nminusacclimit}
  \bar{n}_\infty(x) &= n_0(x) \frac{1-\gamma}{1 + \gamma \cot^2\left(\phi/2 - kx\right)}.
\end{align}
For no loss, $\gamma \to 0$, the modulating factor from the measurement cancels and the original pulse is \emph{completely recovered}, which is surprising since for the first pass there is an anomalously large position shift.  This means that if all the photons in a perfectly collimated pulse are collected through repeated recycling, then the information about the measurement will be erased due to a progressive smearing---or walk-off---effect of the interference pattern, as illustrated in Figs.~\ref{fig:beam1}, \ref{fig:beam2a}, and \ref{fig:beam3a}.  Such a result indicates that a collimated pulse should be thrown away or reshaped after a finite number of traversals in order to maximize the information collected at the detector regarding the momentum kick $k$ and the induced phase shift $\phi$.

\subsubsection{Gaussian Pulse}\label{sec:regimes}
To gain some intuition about the collected number density \eqref{eq:nminusacc}, consider an initial pulse with a zero-mean Gaussian transverse profile.
\begin{align}\label{eq:gaussian}
  n_0(x) = \frac{N}{\sqrt{2\pi\sigma^2}}e^{-x^2/2\sigma^2}.
\end{align}
In what follows, we will consider three specific parameter regimes for the Gaussian pulse:
\begin{enumerate}
  \item the strongly misaligned regime $\phi < 1 < k\sigma$ 
  \item the weak-value regime $k\sigma < \phi < 1$ 
  \item the inverse weak-value regime $\phi < k\sigma < 1$
\end{enumerate}
These regimes are illustrated in Figures~\ref{fig:beam1}, \ref{fig:beam2a}, and \ref{fig:beam3a}, respectively.

In the strongly misaligned regime $\phi < 1 < k\sigma$, the profile that exits the dark port on each traversal \eqref{eq:nminus} is shown in Fig.~\ref{fig:beam1}.  The interference pattern covers the entire beam profile.  On the first pass, the intensity of the peaks match the maximum intensity of the beam.  Subsequent passes are strongly attenuated due to the small overlap with the complementary interference pattern in the beam that remains inside the interferometer.  The accumulated profile $\bar{n}_r(x)$ in \eqref{eq:nminusacc} steadily shrinks the width of the interference dips with increasing traversal number until the entire beam profile is recovered.  The strongly misaligned regime is unlikely to be useful in a precision measurement due to the large value of $k$; we have included it in our discussion for completeness and to emphasize that the single and double lobes that appear in the other regimes are not simple beam shifts, but appear from an interference effect.

For $k\sigma < \phi < 1$, we obtain the weak-value parameter regime considered in \cite{Dixon2009}.  The interference pattern in the number density $n_r(x)$ indicated in \eqref{eq:nminus} leaves a single displaced peak that resembles a shifted Gaussian that is shown in Fig.~\ref{fig:beam2a}.  Subsequent traversals have similar intensities, but progressively walk toward the center with increasing $r$.  The amplified signal comes from the anomalously large shift, so this walk-off degrades the amplification properties of the setup with increasing $r$.

The walk-off effect arises because the beam that remains inside the interferometer has had a small fraction of light removed by the post-selection from one side, which causes a complementary displacement in the opposing direction.  This complementary shift counter-acts the dark-port displacement on subsequent traversals, which makes the output walk back toward the center of the original profile.  Hence, after $r$ traversals the accumulated weak-value signal $\bar{n}_r(x)$ \eqref{eq:nminusacc} will resemble $r$ times the intensity of a single traversal, but will also be degraded due to the walk-off effect.  The walk-off is shown exaggerated in Fig.~\ref{fig:beam2a}, but is a smaller effect per traversal with more realistic parameters, such as those in \cite{Dixon2009}.  However, even though the effect per traversal is smaller, for a sufficiently large number of traversals the signal will always be completely erased by this walk-off effect according to \eqref{eq:nminusacclimit}.

For $\phi < k\sigma < 1$, we enter the inverse weak-value regime considered in \cite{Starling2010a,Starling2010b} and originally observed in \cite{Ritchie1991}.  The weak-value assumptions that produce the single peak break down and \eqref{eq:nminus} produces the double-lobed profile shown in Fig.~\ref{fig:beam3a}.  On multiple traversals the peaks gradually walk back toward the center, similarly to the weak-value regime.  However, the forced zero in the center will stabilize the profile, so that after $r$ traversals the accumulated profile $\bar{n}_r(x)$ \eqref{eq:nminusacc} will more closely resemble $r$ times the intensity of a single traversal than in the weak-value regime.  The walk-off is shown exaggerated in Fig.~\ref{fig:beam3a}, but is also a smaller effect per traversal with more realistic parameters, such as those used in \cite{Starling2010a,Starling2010b}.

These different regimes for weak value amplification measurements are also carefully explored in the recent review paper \cite{Kofman2012}.

\subsubsection{Parity Flips}\label{sec:parity}
A simple technique for compensating for the profile erosion on multiple traversals is to invert the profile around the $x=0$ line so that each new traversal partially cancels the walk-off from the previous traversal.  This can be accomplished by introducing a parity-flipping optic represented by a parity operator $\op{P}_x$ that modifies the profile by replacing $x \to -x$.  This results in a net replacement of the operator $\op{M}_+ \to \op{P}_x \op{M}_+$ in \eqref{eq:measops}.  After an \emph{even} number of traversals $2r$, the accumulated number density \eqref{eq:nminusacc} then has the modified form,
\begin{align}
  \bar{n}_{2r}(x) &= n_0(x) \sin^2\left(\frac{\phi}{2} -kx\right)\left(1+\cos^2\left(\frac{\phi}{2} - kx\right)\right)\times \nonumber \\
  &\quad \frac{1 - \left[\cos^2\left(\frac{\phi}{2} - kx\right)\cos^2\left(\frac{\phi}{2} + kx\right)\right]^r}{1 - \cos^2\left(\frac{\phi}{2} - kx\right)\cos^2\left(\frac{\phi}{2} + kx\right)}, 
\end{align} 
where we have set $\gamma \to 0$ for clarity.  Unlike \eqref{eq:nminusacc} where we do not flip the output of the bright port on each traversal, this expression does not yield the original input profile in the limit of large $r$.  Instead it yields,
\begin{align}
  \bar{n}_{\infty}(x) &= n_0(x) \frac{\sin^2\left(\frac{\phi}{2} -kx\right)\left(1+\cos^2\left(\frac{\phi}{2} - kx\right)\right)}{1 - \cos^2\left(\frac{\phi}{2} - kx\right)\cos^2\left(\frac{\phi}{2} + kx\right)}, 
\end{align} 
which maintains a signal, in contrast with the case of no parity flips on a collimated beam.  

\subsubsection{Zeno stabilization}\label{sec:zeno}
\begin{figure*}[t]
  \begin{center}
    \includegraphics[width=6in]{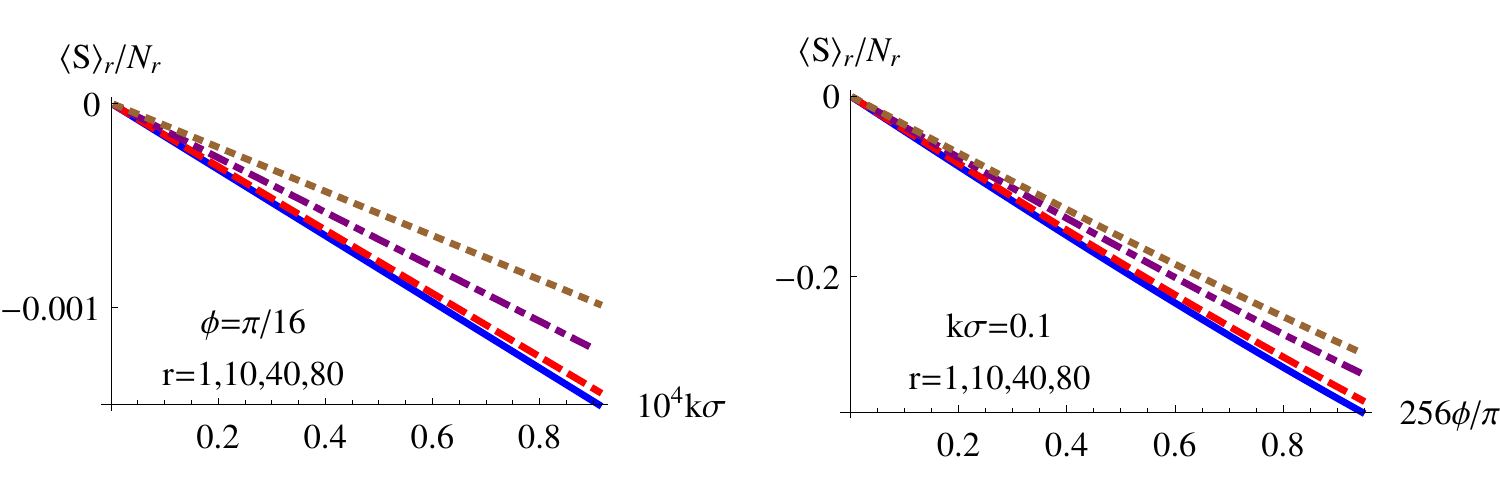}
  \end{center}
  \caption{Normalized split-detection response for a collimated pulse.  Left: response for the weak-value regime as a function of $k\sigma$ with fixed $\phi$, where $k\sigma < \phi < 1$, and with parameters consistent with Ref.~\cite{Dixon2009}.  The accumulated signal is shown for traversal numbers $r=1$ (blue, solid), $r=5$ (red, dashed), $r=10$ (purple, dot-dashed), and $r=20$ (brown, dotted).  Right: response for the inverse weak-value regime as a function of $\phi / \pi$ with fixed $k$, where $\phi < k\sigma < 1$, and with parameters consistent with Ref.~\cite{Starling2010a,Starling2010b}.  The accumulated signal is shown for the same traversal numbers and color coding.  Though the walk-off effects at large traversal numbers change the slope in both regimes, the linear response is preserved.  Hence, one can calibrate the slope through repeated experiments with a fixed number of traversals per laser period.  The slope is negative here since the signal is negative in Eq. \eqref{eq:gaussianpass}.}
  \label{fig:collimatedsplitresponse}
\end{figure*}

\begin{figure*}[t]
  \begin{center}
    \includegraphics[width=6in]{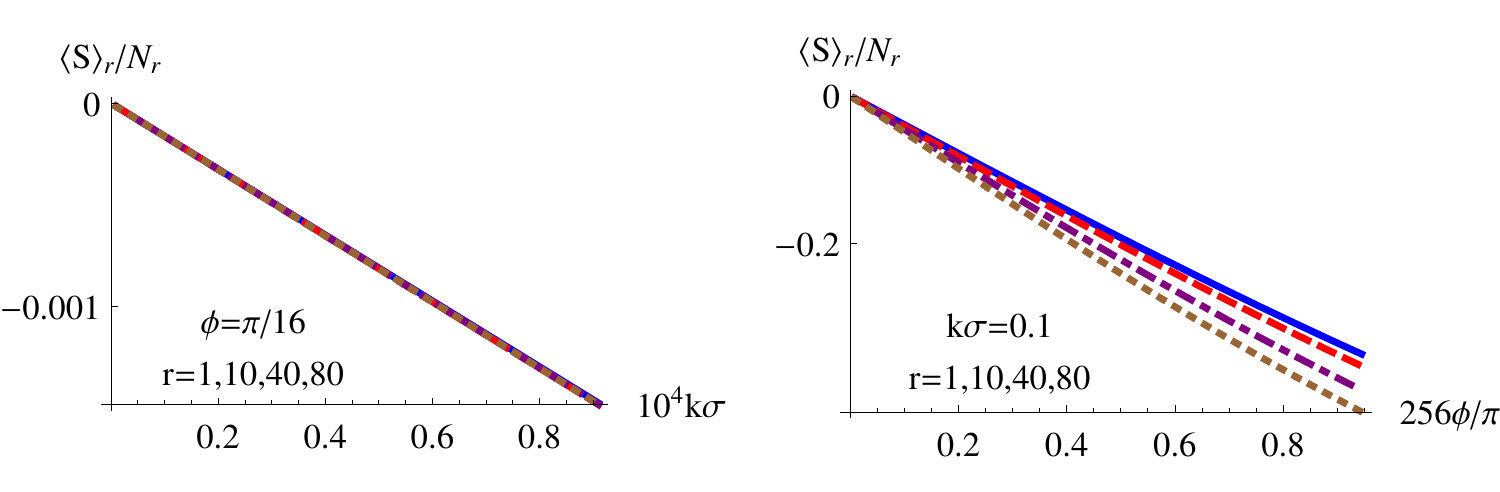}
  \end{center}
  \caption{Normalized split-detection response for a collimated pulse stabilized by parity flipping, using the same color coding as in Figure~\ref{fig:collimatedsplitresponse}.  Left: response for the weak-value regime as a function of $k\sigma$ for fixed $\phi$, where $k\sigma < \phi < 1$.  All traversal numbers $r$ have the same linear response due to the parity-flip stabilization; Zeno stabilization produces the same result.  Right: response for the inverse weak-value regime as a function of $\phi / \pi$ for fixed $k$, where $\phi < k\sigma < 1$.  The linear response acquires a steeper slope for larger traversal number using parity-flip stabilization; Zeno stabilization, however, would result in identical slopes for any $r$. }
  \label{fig:collimatedsplitresponseflip}
\end{figure*}
Another way to reduce transverse walk-off effects and thereby restore the signal-to-noise ratio to the power-limited scaling of \eqref{eq:powergain} in Section~\ref{sec:snr} is to utilize the physics of the quantum Zeno effect by using an optical filter to project the transverse profile back into its original state.  The advantage of the Zeno stabilization over parity flipping is that the former does not swap the transverse locations of the photons; this may be important when using quantum states of light, e.g., squeezed or entangled states, whose benefits rely on maintaining transverse correlations between the photons.

At every round of the recycling with Zeno stabilization, the beam is passed through a spatial filter, so if the beam is in its original profile, it will pass through the filter perfectly; however, if the waveform is distorted, then a photon in that mode will have some probability to be absorbed.  In passing through the filter on each traversal, a photon in this mode will only experience a small disturbance to the transverse profile, and the state will tend to ``freeze'' in its original state with only a small rate of being projected into an orthogonal state (in this case, being absorbed by the filter).  We note this technique will work regardless of the nature of the disturbance, provided it is small in each pass.  

To see how this works, let us consider the Gaussian transverse state in \eqref{eq:gaussian} for a single photon ($N=1$).  The corresponding transverse spatial state for this photon has the form
\begin{align}
  \phi_0(x) = \ipr{x}{\phi_0} = \frac{1}{(2 \pi \sigma^2)^{1/4}} \exp \left( - \frac{x^2}{4 \sigma^2}\right).
\end{align}
After one traversal through the interferometer, the state emerging from the bright port according to \eqref{eq:measops} is  
\begin{align}
  \ket{\phi_1} = \op{M}_+ \ket{\phi_0} = \sqrt{1-\gamma} \cos( \phi/2 - k  \op{x}) \ket{\phi_0}.
\end{align}
To compute the reshaping probability, we renormalize this state by dividing out its norm 
\begin{align}
  \ipr{\phi_1}{\phi_1} &= (1- \gamma) \int \! \textrm{d}x |\phi_0(x)|^2  \cos^2( \phi/2 - k x) \\
  &= (1- \gamma) (1 + e^{-2 k^2 \sigma^2} \cos \phi)/2, \nonumber
\end{align}
to produce the normalized state $\ket{\phi_{1,n}} = \ket{\phi_1} /\sqrt{\ipr{\phi_1}{\phi_1}}$.

If we now make a projective measurement with a spatial filter of the shape $|\ipr{x}{\phi_0}|^2$, the photon will be restored to the state $\ket{\phi_0}$ with a probability $P_1 =  |\ipr{\phi_0}{\phi_{1,n}}|^2$.  The probability can be calculated from
\begin{align}
  \ipr{\phi_0}{\phi_{1,n}} &= \frac{1}{\sqrt{N_1}} \int \! \textrm{d}x |\phi_0(x)|^2 \cos( \phi/2 - k x), \\
  &= \frac{\sqrt{2} e^{-k^2 \sigma^2/2} \cos (\phi/2)}{\sqrt{1 + e^{-2 k^2 \sigma^2} \cos \phi}}. \nonumber
\end{align}
We are interested in the case where both $\phi$ and $k \sigma$ are less than 1.  Consequently, we can expand $P_1$ to leading order in $k \sigma$ and $\phi$,
\begin{align}\label{eq:p1}
  P_1 = 1 - (k \sigma)^4/2 - (k \sigma)^2 \phi^2/4 + \dots,
\end{align}
where we drop terms of higher order in powers of $(k \sigma)^2$ and $\phi^2$.  In the weak-value regime where $k\sigma < \phi < 1$, the second term in \eqref{eq:p1} may be dropped.  In the inverse weak-value regime where $\phi < k\sigma < 1$, the third term in \eqref{eq:p1} may be dropped.  In either case, for repeated cycles consisting of $M$ independent measurements the probability $P_M = P_1^M$ of being projected back into state $\ipr{x}{\phi_0}$ will decay approximately exponentially as 
\begin{align}
  P_M = \exp [ -M \Gamma],
\end{align}
where $\Gamma \approx (k \sigma)^4/2 + (k \sigma)^2 \phi^2/4$ is an effective decay rate.  We can therefore make $M \sim M_Z = 1/\Gamma$ measurements before a photon is typically absorbed by the reshaping filter.  This is the manifestation of the Zeno effect, where by making repeated projections, the state is kept in its initial state for much longer than would happen otherwise.

This Zeno number $M_Z$ is many more cycles that we will be able to make before the detector measures all the photons exiting the dark port.  For example, if we chose the exaggerated values $k \sigma = 0.1$ and $\phi=\pi/8$ as in Figure~\ref{fig:beam2a} then this gives a Zeno number of $M_Z \approx 2.3 \times 10^3$, which is still an order of magnitude larger than we require for the detection physics.

\subsection{Split-detected signal}
\begin{figure*}[th]
  \begin{center}
    \includegraphics[width=6in]{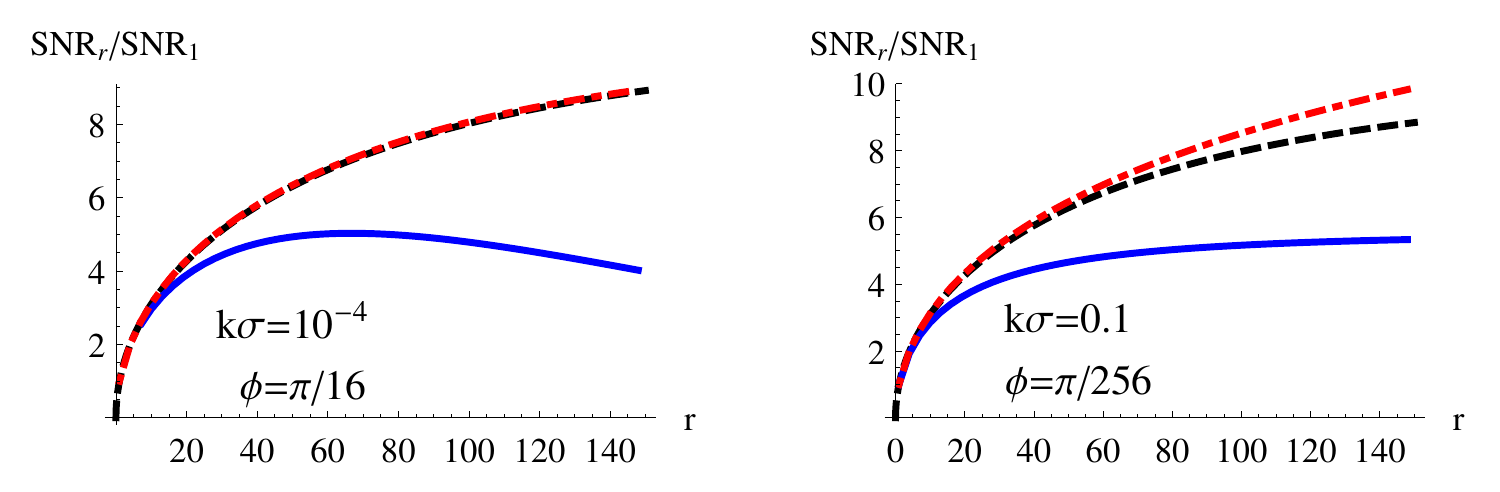}
  \end{center}
  \caption{SNR gain versus traversal number $r$ for a collimated pulse.  Left: the SNR gain for the weak-value regime $k\sigma < \phi < 1$ with $k\sigma = 10^{-4}$ and $\phi = \pi/16$.  The uncorrected beam with walkoff (blue, solid) shows clear degradation with traversal number, while the beam corrected with parity flipping (red, dashed) as in Sec.~\ref{sec:parity} matches the simple power scaling law exactly (black, dashed) from Eq.~\eqref{eq:powergain}. Note that Zeno stabilization will identically produce this power-scaled SNR by construction.  Right: the SNR gain for the inverse weak-value regime $\phi < k\sigma < 1$ with $k\sigma = 0.1$ and $\phi = \pi/256$.  The uncorrected double-lobed beam with walkoff (blue, solid) shows somewhat less degradation than the weak-value regime due to the forced zero in the profile; however, the double-lobed beam corrected with parity flipping (red, dashed) manages to exceed the simple power scaling law (black, dashed), and thus the scheme with Zeno stabilization.}
  \label{fig:collimatedsnr}
\end{figure*}

In order to measure the transverse momentum kick $k$ or the phase shift $\phi$ we compare the sides of the transverse profile using a split detector.  As outlined in Section \ref{sec:snr}, the accumulated split-detected signal after $r$ pulse repetitions of time duration $\tau$ is given by the difference of the number densities on each side,
\begin{align}
  \mean{S}_r &= \int_0^\infty \!\textrm{d}x\, \bar{n}_r(x) - \int_{-\infty}^0 \!\textrm{d}x\, \bar{n}_r(x).
\end{align}
To measure the displacement of the pulse, the signal should be subsequently normalized by the total photon number $\mean{S}_r / N_r$ in order to extract the averaged behavior.

For small displacements the variance of the raw split detected signal is the second moment to a good approximation,
\begin{align}
  (\Delta S)^2_r \approx \mean{S^2}_r &= \int \! \textrm{d}x \, \bar{n}_r(x) = N_r,
\end{align}
which is the total number of photons that have impacted at the dark port detector.  Hence, the SNR that has accumulated after the $r$\textsuperscript{th} traversal will be given by,
\begin{align}
  \text{SNR}_r &= \frac{\mean{S}_r}{(\Delta S)_r} = \frac{\mean{S}_r}{\sqrt{N_r}}.
\end{align}

For the zero-mean Gaussian \eqref{eq:gaussian} these quantities can be computed exactly for the first traversal, 
\begin{subequations}\label{eq:gaussianpass}
\begin{align}
  N_1 &= \frac{(1-\gamma) N}{2}\left(1 - e^{-(2k\sigma)^2/2}\cos\phi\right), \\
  \mean{S}_1 &= - \frac{(1-\gamma) N}{2} e^{-(2k\sigma)^2/2} \text{Erfi}(\sqrt{2}k\sigma) \sin\phi, \displaybreak[0]\\
  \frac{\mean{S}_1}{N_1} &= -\text{Erfi}(\sqrt{2}k\sigma) \frac{e^{-(2k\sigma)^2/2}\sin\phi}{1 - e^{-(2k\sigma)^2/2}\cos\phi}, \\
  \text{SNR}_1 &= - \sqrt{(1-\gamma)N}\, \frac{\text{Erfi}(\sqrt{2}k\sigma) e^{-(2k\sigma)^2/2}\sin\phi}{\sqrt{2(1 - e^{-(2k\sigma)^2/2}\cos\phi)}},
\end{align}
\end{subequations}
where $\text{Erfi}(x) = \text{Erf}(ix)/i = (2/\sqrt{\pi})\int_0^x\! e^{t^2} \mathrm{d}t$ is the imaginary error function.  We now specialize these exact solutions to the two amplification regimes under consideration and indicate numerically how larger traversal numbers behave in each regime.

\subsubsection{Weak-value regime}
When $k\sigma < \phi < 1$ then we can neglect terms of order $(k\sigma)^2$ in \eqref{eq:gaussianpass} to find,
\begin{subequations}\label{eq:splitgausslinear}
\begin{align}
  N_1 &= (1-\gamma) N \sin^2(\phi/2), \\
  \mean{S}_1 &= - \sqrt{\frac{2}{\pi}} (1-\gamma) N k\sigma \sin\phi, \displaybreak[0]\\
  \frac{\mean{S}_1}{N_1} &= - \sqrt{\frac{2}{\pi}} 2 k \sigma \cot(\phi/2), \\
  \text{SNR}_1 &= - \sqrt{\frac{2}{\pi}} \sqrt{(1-\gamma)N} (2 k\sigma \cos(\phi/2)) \\
  &= - \sqrt{\frac{2}{\pi}} \sqrt{N_1} (2 k\sigma \cot(\phi/2)). \nonumber
\end{align}
\end{subequations}
These linear order solutions correctly match the weak value analyses made in \cite{Dixon2009,Starling2009,Howell2010}, as expected.  Due to the factor $\cot(\phi/2)$ in the normalized split detection $\mean{S}_1/N_1$, setting a known small $\phi$ provides an amplification factor for measuring an unknown small $k$.  This regime gets its name from the fact that this amplification factor is the imaginary part of the weak value $W_w = \bra{\psi_-}\op{W}\ket{\psi_\phi}/\ipr{\psi_-}{\psi_\phi} = i \cot(\phi/2)$ of the which-path operator $\op{W}$ with initial state $\ket{\psi_\phi} = \op{U}_{\text{SBC}}\ket{\psi_+}$ and post-selection state $\ket{\psi_-}$.  The normalized signal for this parameter regime is shown in the left plot of Fig.~\ref{fig:collimatedsplitresponse} as a function of $k\sigma$, demonstrating the linear response.

We can reproduce the dominant SNR gain factor for small post-selection probability by neglecting the walk-off effects and the power attenuation.  To do this, we expand the accumulate profile $\bar{n}_r(x)$ in \eqref{eq:nminusacc} to first order in $k\sigma$ and second order in $\phi$ to obtain,
\begin{align}\label{eq:wvn}
  \bar{n}_r(x) &= r c(\gamma,r) n_0(x) \left(- k x \phi + \left(\frac{\phi}{2}\right)^2\right), \\
  \label{eq:wvc}
  c(\gamma,r) &= \frac{(1-\gamma)(1 - (1-\gamma)^r)}{r \gamma},
\end{align}
where $\lim_{\gamma\to 0}c(\gamma,r) = 1$.

The only $r$-dependence in the number density is in the numeric prefactor $r c(\gamma,r)$, which effectively scales the total photon number $N \to r c(\gamma,r) N$.  Using this scaling, the result \eqref{eq:splitgausslinear} for the split-detector will hold for any $r$ to second order in $\phi$ and first order in $k$.  Hence, the SNR should scale as $\sqrt{N r c(\gamma, r)}$ when walk-off and power attenuation effects are neglected.  When $\gamma \to 0$, this recovers the dominant $\sqrt{r}$ SNR enhancement factor that we found to zeroth order in the post-selection probability of \eqref{eq:powergain} in Section \ref{sec:snr} from power considerations.

However, the walk-off effects and power attenuation combine to reduce the actual SNR below this optimistic level.  To see this, consider the solid blue curve in the left plot of Fig.~\ref{fig:collimatedsnr}, which shows the split-detected SNR gain versus traversal number for the weak-value regime.  The SNR gain for any sufficiently small $k\sigma$ is universal, but plateaus quickly due to the beam degradation from the walk-off.  Even worse, for sufficiently large traversal number $r$ the signal will eventually decline and then converge to zero due to the erasure effect implied by \eqref{eq:nminusacclimit}, so the SNR gain factor will also correspondingly decay to zero.  

The power scaling in \eqref{eq:powergain} can be recovered, however, if the walk-off is corrected with the parity flipping method discussed in Section~\ref{sec:parity}.  The normalized signal produced with the parity-flip correction---illustrated in the left plot of Fig.~\ref{fig:collimatedsplitresponseflip}---has an identical slope for any traversal number, demonstrating the simple power scaling behavior.  This correction is illustrated as the dot-dashed red curve  in the left plot of Fig.~\ref{fig:collimatedsnr}, which exactly overlaps the power scaling curve illustrated as the dashed black curve.  If the walk-off is corrected with Zeno stabilization instead, then the signal slope will be identical for any traversal number by construction, and the SNR gain will also exactly follow the power scaling curve in \eqref{eq:powergain}.

\subsubsection{Inverse weak-value regime}
If $\phi < k\sigma < 1$, then the approximation to linear order in $k\sigma$ will break down, as shown earlier in Section \ref{sec:regimes}.  For this regime, we keep linear order in $\phi$ and second order in $k\sigma$ in \eqref{eq:gaussianpass} to find,
\begin{subequations}\label{eq:splitgaussquad}
\begin{align}
  N_1 &= (1-\gamma) N (k\sigma)^2, \\
  \mean{S}_1 &= - \sqrt{\frac{2}{\pi}} (1-\gamma) N k \sigma \phi, \displaybreak[0]\\
  \frac{\mean{S}_1}{N_1} &= \sqrt{\frac{2}{\pi}} \left(\frac{k\sigma}{3}  - \frac{1}{k\sigma} \right) \phi, \\
  \text{SNR}_1 &= \sqrt{\frac{2}{\pi}} \sqrt{(1-\gamma)N} \left(\frac{5}{6}(k\sigma)^2 - 1\right)\phi \\
  &= \sqrt{\frac{2}{\pi}} \sqrt{N_1}\left(\frac{5}{6} k\sigma - \frac{1}{k\sigma}\right)\phi. \nonumber
\end{align}
\end{subequations}
In contrast to the previous approximation, the $1/k$ term in the normalized signal $\mean{S}_1/N_1$ leads to an amplification in measuring an unknown $\phi$ given a known small $k$.  Indeed, this regime was used in Refs.~\cite{Starling2010a,Starling2010b} for exactly this purpose.  In the preprint version of Ref.~\cite{Starling2010a} it was noted that $\phi \approx 2\text{Im}W_w^{-1}$ is the inverse of the weak value present in the weak-value regime for small $\phi$, which motivates our name for this parameter regime; this inverted relationship has also been rediscovered more recently in Ref.~\cite{Kofman2012}.  The normalized signal for this parameter regime is shown in the right plot of Fig.~\ref{fig:collimatedsplitresponse} as a function of $\phi$, demonstrating the linear response.

Again, we can reproduce the dominant SNR gain factor for small post-selection probability by neglecting the walk-off effects and the power attenuation, which can be done by expanding $\bar{n}_r(x)$ to second order in $k\sigma$ and first order in $\phi$ to obtain,
\begin{align}
  \bar{n}_r(x) &= r c(\gamma,r) n_0(x) \left(- k x \phi + k^2 x^2 \right),
\end{align}
with the same $c(\gamma,r)$ as in \eqref{eq:wvc}.

As with the weak-value regime, the only $r$-dependence in the number density is in the numeric prefactor $r c(\gamma,r)$, which effectively scales the total photon number $N \to r c(\gamma,r) N$.  Using this scaling, the result \eqref{eq:splitgaussquad} for the split-detector will hold for any $r$ to second order in $k$ and first order in $\phi$.  Hence, the SNR will scale as $\sqrt{N r c(\gamma, r)}$ when walk-off and power attenuation effects are neglected.  When $\gamma \to 0$, this also recovers the dominant $\sqrt{r}$ SNR enhancement factor that we found to zeroth order in the post-selection probability of \eqref{eq:powergain} in Section \ref{sec:snr} from power considerations.

As before, the walk-off and power attenuation effects reduce the SNR gain below this optimistic level.  Consider the solid blue curve in the right plot of Fig.~\ref{fig:collimatedsnr}, which shows the exact split-detected SNR gain versus traversal number for the inverse weak-value regime.  As anticipated in Section \ref{sec:regimes}, the forced zero in the center of the double-lobed profile naturally stabilizes the beam to produce a saturated SNR for more traversals than the weak-value regime.  However, the SNR still plateaus relatively quickly before eventually decaying to zero for a sufficiently large number of traversals $r$ without additional stabilization.  

Similarly, the degradation from walk-off can be completely reversed by employing the parity flipping technique.  The dot-dashed red curve in the right plot of Fig.~\ref{fig:collimatedsnr} actually \emph{exceeds} the simple power scaling law illustrated as the dashed black curve due to an additional accumulation of momentum information on each subsequent traversal.  Moreover, the normalized signal shown in the right plot of Fig.~\ref{fig:collimatedsplitresponseflip} shows a corresponding increase in the slope of SNR vs. $\phi$ with traversal number.

\begin{table}[t]
  \centering
  \begin{tabular}{c | c}
    Symbol \; & \; Numerical Value \\
    \hline
    $\ell$ & \; 1.5 m \\
    $\sigma$ & \; $1$ mm \\
    $k$ & \; $1 \times 10^{-3} $ m\textsuperscript{-1}\\
    $k_0$ & \; $8 \times 10^6 $ m\textsuperscript{-1}\\
    $\gamma$ & \; 0.01\\
    $d$ & \; $1$ cm
  \end{tabular}
  \caption{Parameters used for numerical computations.  $\ell$ is half the length of the interferometer, $k$ is the momentum kick from the mirror, $\sigma$ is the input beam width, $k_0$ is the carrier momentum, $\gamma$ is the loss per traversal, and $d$ is the half-width of the split-detector.}
  \label{tab:numericalparameters}
\end{table}

\section{Diverging Pulse}\label{sec:diverging}
\begin{figure}[t]
  \includegraphics[width=3in]{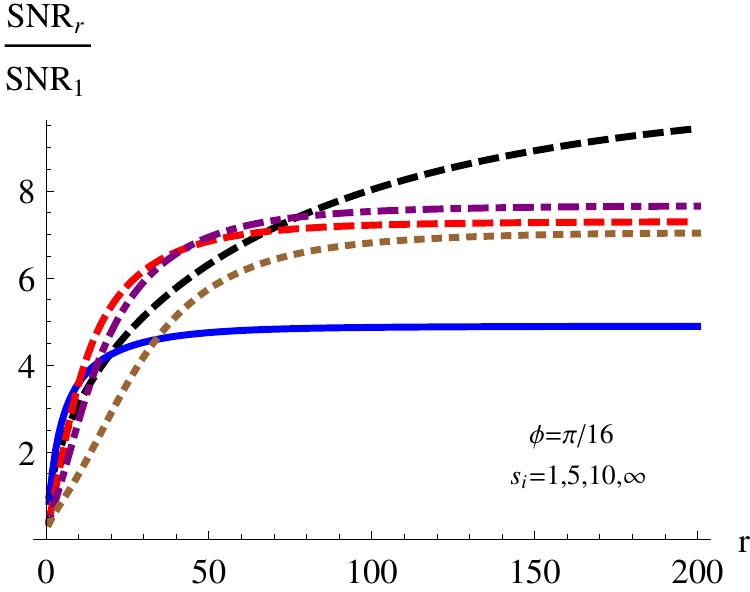}
  \caption{The effect of propagation and different diverging lens choices on the SNR of the weak-value regime with $k\sigma = 10^{-6}$ and $\phi = \pi/16$.  In order of (blue, solid), (red, dashed), (purple, dot-dashed), and (brown, dotted) we show weak initial diverging lenses with extreme focal lengths $s_i = -1\text{ m,} -5\text{ m,} -10\text{ m,}$ and no lens.  The (black, dashed) curve shows the scaling given by power considerations in Eq.~\eqref{eq:powergain}. Here $\text{SNR}_1$ refers to the SNR of a single unrecycled pulse with an optimally chosen focal length of $s_i = -0.5\text{ m}$, while $\text{SNR}_r$ is the accumulated SNR over $r$ traversals for the indicated lens choices. }  
  \label{fig:snrplateaus}
\end{figure}
By assuming a collimated beam, we have so far neglected beam propagation effects in the analysis, as well as any lens effects that could further change the detection physics.  In order to incorporate these effects, we now alter our measurement operators and pursue a numerical approach.  We find that these effects may slightly enhance the SNR gains from recycling before saturation due to the finite detector size, but do not fundamentally alter the basic power-scaling behavior.  For that reason, it will be sufficient to illustrate only the weak-value regime as an example.

Extending the collimated analysis in Section~\ref{sec:collimated} to include beam propagation leads to a replacement of the measurement operators with $\op{M}_\pm \to \op{M}'_\pm = \op{U}_\ell \op{M}_\pm \op{U}_\ell$, where $\op{U}_\ell = \exp(-i \op{p}^2 \ell / 2 k_0)$, $k_0$ is the carrier momentum of the pulse, and $\ell$ is the propagation length from piezo to 50:50 beam splitter \cite{Born1959}.  The number density \eqref{eq:nminus} will then involve the composite measurement operator 
\begin{align}\label{eq:recursivetotal}
  \op{M}_-^{'}\left(\op{M}_+^{'}\right)^{r-1} &= \op{U}_\ell\op{M}_-\op{U}_\ell \left(\op{U}_\ell\op{M}_+\op{U}_\ell\right)^{r-1}.
\end{align}  
Adding a diverging lens with focal length $s_i$ also modifies the initial state with an operator $\op{U}_L = \exp(i k_0 \op{x}^2 / 2 s_i)$.  Hence, powers of the following modified operators will appear in the full solution,
\begin{subequations}\label{eq:recursiveparts}
\begin{align}
  \op{U}_\ell\op{M}_- &= i e^{-i\op{p}^2\ell/2k_0}\sin\left(\phi/2 - k\op{x}\right) , \\
  \op{U}_\ell^2 \op{M}_+ &= e^{-i\op{p}^2\ell/k_0}\cos\left(\phi/2 - k\op{x}\right) , \\
  \op{U}_\ell\op{U}_L &= e^{-i\op{p}^2\ell/2k_0}e^{ik_0\op{x}^2/2s_i},
\end{align}
\end{subequations}
which can be simplified recursively, as detailed in Appendix~\ref{sec:recursiveappendix}.  The effect of a diverging lens is considered for comparison with the unrecycled experiment in Ref.~\cite{Dixon2009}, where such a lens was able to enhance sensitivity.  

Table~\ref{tab:numericalparameters} shows the parameters which describe the laser and experimental geometry. Our choice of $k_0$ corresponds to the 780-800 nm lasers used in \cite{Dixon2009,Starling2009,Starling2010a,Starling2010b}, and the 3-meter interferometer length $\ell$ is taken from the generous upper bound estimate discussed in Section~\ref{sec:qual} as a worst case scenario for beam divergence effects.  For the weak-value regime of small $k\sigma$ and $\phi$ such that $k\sigma < \phi < 1$, we found that it was more computationally efficient to expand the sine and cosine functions in \eqref{eq:recursiveparts} to second order in $k$ and fourth order in $\phi$, as shown in Appendix~\ref{sec:truncation}.  To test the validity of this truncation, we initially set the interferometer length $\ell$ to zero so that a comparison could be made with the previously calculated collimated solutions.

We restrict our attention to the SNR gains achieved by recycling a single pulse for $r$ traversals, since adding more pulses leads to a simple scaling of the single pulse result. The SNR gains for different choices of initial diverging lens are shown in Fig.~\ref{fig:snrplateaus}, where they are compared to the ideal power-scaling curve that we expect from our qualitative considerations given by \eqref{eq:powergain}.  In all cases, the expected gains roughly follow the qualitative power scaling rule for a large number of traversals before saturating due to the beam growing larger than the finite size of the split-detector.  Note that the beam divergence mitigates the SNR decay that was observed for the collimated case, even without beam stabilization due to flipping or Zeno reshaping.

\section{Conclusion}\label{sec:conclusion}
By investigating the optical design shown in Fig.~\ref{fig:setup}, we have shown how a single optical pulse can be trapped inside the interferometer until the photons all exit the dark port and are ``post-selected,'' greatly boosting the sensitivity of the precision measurement.  The added power accumulated at the detector within a fixed duration of time is the dominant source of sensitivity gain.  Further increases are achievable by trapping multiple pulses in the interferometer simultaneously.  The number of trapped pulses is limited by the length of the pulses, the gating frequency of current Pockels cells, and the physical size of the interferometer.

We carefully analyzed the case of a collimated beam and showed that repeated post-selections cause a walk-off effect in the recycled pulse, which tends to diminish the SNR.  However, we also showed that these walk-off effects can be easily corrected by Zeno reshaping, or by a parity flip, which reflects the beam around its optic axis on each traversal.  Somewhat surprisingly, the gains with parity correction can even exceed those expected from the power scaling.  Including propagation effects does not destroy the sensitivity gain shown for the collimated case, but instead can produce additional enhancement.  

While these sensitivity gains alone are a substantial improvement over the original idea, the combination of these techniques with other established metrology techniques---such as the use of a squeezed reference beam---could further increase the sensitivity beyond that indicated here.

\begin{acknowledgments}
  We acknowledge support from the US Army Research Office Grants No. 62270PHII: STIR, and No. W911NF-09-0-01417, as well as the National Science Foundation Grant No. DMR-0844899.
\end{acknowledgments}

%

\appendix
\section{Recursive Simplification}\label{sec:recursiveappendix}
Using Equations \eqref{eq:recursivetotal} and \eqref{eq:recursiveparts} from above, an exact recursive simplification can be constructed.  Written explicitly in the momentum basis we have 
\begin{align}
\nonumber\op{U}_\ell\op{M}_-\left(\op{U}_\ell^2 \op{M}_+\right)^{r-1} &= \left(\frac{1}{2}\right)^{r}\int\! \mathrm{d}p\ket{p}e^{-i\ell p^2/2k_0} \times \\ &\sum^r_{j=-r}a_j(r,p) e^{ij\phi/2}\bra{p + jk},
\end{align} 
where the functions $a_j(r,p)$ are given by the recursion relation
\begin{align}
\nonumber a_{j}(r,p) = a_{j-1}(r-1,p)e^{-i\ell(p+(j-1)k)^2/k_0} +\\ a_{j+1}(r-1,p)e^{-i\ell(p+(j+1)k)^2/k_0}.
\end{align}
The overall measurement operator is then given simply by multiplying the unitary operators $\op{U}_\ell$ and $\op{U}_L$ from the right.  Note that each pulse retains the phase and momentum information accumulated from previous traversals.   

Given an input Gaussian profile as in \eqref{eq:gaussian}, we find 
\begin{align}\label{eq:exactsolution}
\ipr{x}{\Psi_-^r} =& \left(\frac{1}{2}\right)^{r}\left(\frac{2}{\pi a^2}\right)^{1/4} \beta \times \\
&\quad \int \!\mathrm{d}p e^{ipx} e^{-i\ell p^2/2k_0}e^{-i\ell (p+k)^2/k_0}  \times \nonumber \\
&\quad \sum_{j=-r}^{r}a_j(r,p)e^{ij\phi/2}e^{-i\ell(p+jk)^2/2k_0}e^{-\beta^2(p+jk)^2}, \nonumber 
\end{align}
with 
\begin{align}
\beta = \sqrt{\frac{\sigma^2 s_i}{s_i - 2ik_0\sigma^2}}.
\end{align} 

The solution with no initial diverging lens is given by taking limit $s_i \rightarrow \infty$, which modifies Equation~\eqref{eq:exactsolution} with the replacement $\beta \rightarrow \sigma$.  Similarly, setting $\ell = 0$ recovers the collimated solution as expected. 

\section{Numerical Truncation}\label{sec:truncation}
Truncating the measurement operators that include propagation effects to second order in $k$ and fourth order in $\phi$ for the weak value regime produces the expressions, 
\begin{subequations}\label{eq:approximateexpansions}
\begin{align}
 \op{U}_\ell\op{M}_- &= \int \! \mathrm{d}p\ket{p}e^{-ip^2\ell/2k_0}\, \times \\
 &\qquad \left[\frac{i\phi}{2} -\frac{i\phi^3}{48} + k\partial_p - \frac{k\phi^2}{8}\partial_p + \frac{ik^2\phi}{4} \partial_p^2\right]\bra{p}, \nonumber \\
 \op{U}_\ell^2 \op{M}_+ &= \int \! \mathrm{d}p\ket{p}e^{-ip^2\ell/k_0}\, \times \\
 &\qquad \left[1 - \frac{\phi^2}{8} + \frac{\phi^4}{384} + \frac{ik\phi}{2}\partial_p - \frac{ik\phi^3}{48}\partial_p \right.\nonumber \\
 &\qquad\quad\left. + \frac{k^2}{2}\partial_p^2 - \frac{k^2\phi^2}{16}\partial_p^2\right]\bra{p} . \nonumber
\end{align}
\end{subequations}
These expansions can be numerically iterated more easily than the full solution, using the initial Gaussian profile \eqref{eq:gaussian} as an input.  

\end{document}